\documentstyle{elsart}
\begin{document}

\begin{flushright}
hep-th/9810245
\end{flushright}

\begin{frontmatter}
\title{$T$--Duality and Spinning Solutions in $2+1$ Gravity}

\thanks[cmc]{Email: chen@grg1.phys.msu.su}

\author{Chiang-Mei Chen\thanksref{cmc}}

\address{Department of Theoretical Physics,
         Moscow State University, \\ 119899, Moscow, Russia}

\begin{abstract}
Starting with the $2+1$ Einstein--Maxwell--Dilaton system with a
cosmological constant and assuming two commuting Killing symmetries
we derive the corresponding $1+0 \; \sigma$--model.
It is shown that, for general values of the coupling parameters, the
$T$--duality group is $SL(2,R)$, which coincides with the group of
linear coordinate transformations along the Killing orbits.
This duality, along with a suitable parameter choice, is applied to
obtain some new spinning solutions in the alternative gravity theories:
Einstein--Maxwell, Brans--Dicke and Einstein--Maxwell--Dilaton.
\end{abstract}

\begin{keyword}
$T$-duality, spinning solutions
\end{keyword}
\end{frontmatter}

\section{Introduction}
Gravity in $2+1$ dimensions has attracted much attention since
Ba\~nados, Teitelboim and Zanelli \cite{BaTeZa92,BaHeTeZa93}
discovered a black hole solution in the ($2+1$)--dimensional Einstein
theory with a negative cosmological constant.
The BTZ black hole is a surprisingly simple solution of $2+1$ gravity
exhibiting almost all usual features of a black hole in spite of the
fact that the spacetime curvature is constant.
Soon after, various spinning electrically charged solutions were found
\cite{Cl93} in the Einstein--Maxwell theory.
Moreover, from an (anti-) self-duality assumption for the Maxwell
field a horizonless regular particle-like ``dyon'' solution was
obtained \cite{KaKo95}.
It has been pointed out that the angular momentum and the mass of the
(anti-) self dual solutions diverge at spatial infinity \cite{Ch96}.

Generally in three dimensions (unlike the four--dimensional case)
there is no duality between electric and magnetic fields except for
the static and rotationally symmetric configurations \cite{KiPa97}.
Thus we should consider the electric and magnetic solutions separately.
The static magnetically charged solution in the Einstein--Maxwell theory
was suggested in \cite{HiWe96} and interpreted as a magnetic monopole;
it has no event horizon and is particle--like.
Finally, using a coordinate transformation, Cl\'ement found general
spinning electrically charged BTZ black holes \cite{Cl96}.
All charged solutions appear to have a logarithmic divergence of mass
and angular momentum densities at spatial infinity.
In order to regularize divergence a topological Chern-Simons term was
added directly \cite{Cl96} or through the boundary conditions
\cite{KaKo97}.

The extended theories including a dilaton field, such as Brans--Dicke and
Einstein--Maxwell--Dilaton (EMD) theories, were studied by many authors.
For Brans--Dicke theory static and stationary black hole solutions were
analyzed in \cite{SaKlLe96,SaLe98} using the string frame.
For EMD theory, Chan and Mann \cite{ChMa94} have found static electrically
charged solutions and later spinning uncharged solutions \cite{ChMa96}.
Similarly to the situation in Einstein--Maxwell theory, magnetic EMD
solutions were also obtained in \cite{KoMaNa97}.
Most of the solutions were found by solving the field equations directly
using different assumptions and technical tricks in order to simplify them.
But no general spinning charged solutions have been found in this way since
the spin necessitates both electric and magnetic fields being present thus
increasing the complexity of the field equations.

Here we suggest to use $T$--duality for generating new solutions.
Assuming that the $2+1$ spacetime permits two Killing vector fields,
we construct an $1+0$ equivalent $\sigma$--model by a dimensional
reduction of the metric and a vector field.
The resulting $\sigma$--model possesses $T$--duality symmetry which can
be directly applied to known solutions to construct some new ones.
In this paper we consider the action including the dilaton and the Maxwell
field with arbitrary coupling constants.
In the general case, the $T$--duality {\em locally} is just a general
coordinate transformation and part of it was used to get stationary
solutions previously \cite{Cl96,SaLe98}.
Here we want to emphasize that {\em globally} the coordinate
transformations used to generate spinning solutions from static ones have
more appropriate interpretation as $T$--duality of the reduced $2+1$ theory.
Moreover, for particular set of coupling constants the symmetry is enhanced
and the system becomes fully integrable.

Throughout this paper, the $T$--duality as well as a suitable choice of
parameters are used to generate spinning solutions.
In the Einstein--Maxwell theory we found a spinning electric solution
which is equivalent to Cl\'ement's solution \cite{Cl96} and a new spinning
magnetic solution.
Moreover a dyon solution can be obtained either from electric or magnetic
one by identifying the spin parameter with the charge parameter.
This is a generalization of the (anti-) self dual solutions \cite{KaKo95}.

For the Brans--Dicke case we had found a set of static and stationary
solutions which are similar to \cite{SaKlLe96,SaLe98} but are obtained
in the Einstein frame.
For the EMD case we found two new classes of spinning solutions: electric
and magnetic.
The spinning charged solution has some special features distinguishing
it from all previous ones.
First, it necessarily contains the region where the angle $\theta$ becomes
time--like and hence there exists closed time--like curves.
Second, the solution possesses an inner horizon.
Unfortunately, we did not find an explicit formula for the radius of
the outer horizon.
Hence the general discussion of the black hole thermodynamics is still
lacking.

\section{$1+0 \; \sigma$--Model and $T$--Duality}
Consider a ($2+1$)--dimensional gravity theory including the Maxwell
and dilaton fields with general coupling parameters.
In the Einstein frame the action reads
\begin{equation}
S = \frac1{2\kappa^2} \int \, d^3x \sqrt{-g} \left\{ R
  - 4 \gamma (\partial \phi)^2 - e^{-4\alpha\phi} F^2
  - 2 e^{\beta\phi} \Lambda \right\},
    \label{action}
\end{equation}
where $R$ is the three--dimensional scalar curvature, $\Lambda$
is a cosmological constant ($\Lambda <0$ corresponds to anti--de--Sitter),
$F$ is the Maxwell two--form field and $\phi$ is the dilaton
which couples to $F^2$ and $\Lambda$ with strengths $\alpha$ and $\beta$
respectively. As particular cases this model includes
Einstein--Maxwell ($\alpha=\beta=\gamma=0$),
Brans--Dicke ($\beta=4$ and $F=0$),
and low energy string theory ($\alpha=1, \beta=4$ and $\gamma=1$).

Field equations corresponding to this action are
\begin{eqnarray}
&& R_{\mu\nu}
   = 4 \gamma \partial_\mu \phi \partial_\nu \phi
   + e^{-4\alpha\phi} \left( 2F_{\mu\lambda} F_\nu{}^\lambda
     - g_{\mu\nu} F^2 \right)
   + 2g_{\mu\nu}e^{\beta\phi} \Lambda, \\
&& 4 \gamma \nabla^2 \phi + 2\alpha e^{-4\alpha\phi} F^2
   - \beta e^{\beta\phi} \Lambda = 0, \\
&& \nabla_\mu \left( e^{-4\alpha\phi} F^{\mu\nu} \right) = 0,
\end{eqnarray}
while the Bianchi identity for the Maxwell field is
\begin{equation}
\epsilon^{\lambda\mu\nu} \partial_\lambda F_{\mu\nu} = 0.
\end{equation}

Now we assume the existence of two commuting Killing vectors, so that
in the adapted coordinates all quantities depend only on one variable
(assumed to be spacelike), which is commonly denoted as $r$:
\begin{equation}
ds^2 = h_{ij}(r) dx^i dx^j + e^\varphi f^{-2}(r) dr^2, \qquad
A = A_i(r) dx^i,
\label{ansatz}
\end{equation}
where $\varphi = \ln |\det h|$. The `radial' component of the
$3$--potential $A_r$ was set zero as a pure gauge quantity.
Substituting this ans\"atz into the action (\ref{action}) and dividing
it by the two--dimensional volume of the Killing orbits one arrives at the
$0+1$ action
\begin{eqnarray}
S = \frac1{2\kappa^2} \int \, dr \Big\{ f \Big(
  &&  \frac14 \partial_r \varphi \partial_r \varphi
  + \frac14 \partial_r h_{ij} \partial_r h^{ij}
  - 4 \gamma \partial_r \phi \partial_r \phi
    \nonumber \\
 && - 2 e^{-4\alpha\phi} h^{ij} \partial_r A_i \partial_r A_j \Big)
  - 2 f^{-1} e^{\beta\phi+\varphi} \Lambda \Big\}.
\end{eqnarray}
Clearly, the function $f$ is pure gauge, and
it enters into the action as a Lagrange multiplyer.

Now it is convenient to fully decouple the scale degree of freedom
normalizing the moduli $h_{ij}$ as
\begin{equation}
H_{ij} = e^{-\varphi/2} h_{ij}, \qquad
H^{-1}_{ij} = e^{\varphi/2} h^{ij}.
\end{equation}
Then, in matrix notation, the action will read:
\begin{eqnarray}
S = \frac1{2\kappa^2} \int \, dr \Big\{ f \Big(
  &&  \frac18 \varphi'^2 + \frac14 Tr( H'H'^{-1})
  - 4 \gamma \phi'^2 \nonumber \\
 && - 2 e^{-4\alpha\phi-\varphi/2} Tr( A'^T H^{-1}A') \Big)
  - 2 f^{-1} e^{\beta\phi+\varphi} \Lambda \Big\},
  \label{Asigma1}
\end{eqnarray}
where $A$ is a column of covariant components $A_{i}$.

It is easy to see that the $\sigma$--model action (\ref{Asigma1}) is invariant
under the $T$--duality transformation:
\begin{equation}
H \to \Omega^T H \Omega, \qquad
  A \to \Omega^T A, \qquad
  \hbox{where} \qquad \Omega \in SL(2,R).
\end{equation}
This transformation holds for any values of $\alpha, \beta$ and $\gamma$
therefore it can be applied to all theories included in the action
(\ref{action}).
It is not a surprise because, locally, this $T$--duality is just a
coordinate transformation.

For practical applications we can parameterize $\Omega$ as
\begin{eqnarray}
\Omega &=& \left( \begin{array}{cc} 1 & 0 \\ \mu & 1 \end{array} \right)
         \left( \begin{array}{cc} \xi & 0 \\ 0 & \xi^{-1} \end{array} \right)
         \left( \begin{array}{cc} 1 & \nu \\ 0 & 1 \end{array} \right)
       = \left( \begin{array}{cc}
                \xi & \xi\nu \\ \xi\mu & \, \xi^{-1}+\xi\mu\nu
                  \end{array} \right), \nonumber \\
\Omega^{-1} &=& \left( \begin{array}{cc}
                     \xi^{-1}+\xi\mu\nu & \, -\xi\nu \\ -\xi\mu & \xi
                     \end{array} \right).
\end{eqnarray}
Assuming the seed solution to have the form
\begin{equation}
h^0_{ij} = \left( \begin{array}{cc} -H_0^2(r) & 0 \\ 0 & H_1^2(r)
                  \end{array} \right), \qquad
A^0_i = \left( \begin{array}{c} -A_0(r) \\ A_1(r)
               \end{array} \right),
    \label{seed}
\end{equation}
after $T$--duality transformation we get a new solution
\begin{eqnarray}
h_{ij} &=& \left( \begin{array}{cc}
                  -\xi^2 H_0^2 + \xi^2\mu^2 H_1^2 &
                  -\xi^2\nu H_0^2+\mu(1+\xi^2\mu\nu) H_1^2 \\
                  -\xi^2\nu H_0^2+\mu(1+\xi^2\mu\nu) H_1^2 &
                  \, -\xi^2\nu^2 H_0^2+\xi^{-2}(1+\xi^2\mu\nu)^2 H_1^2
                  \end{array} \right), \nonumber \\
A_i &=& \left( \begin{array}{c}
               -\xi A_0 + \xi \mu A_1 \\
               -\xi \nu A_0 + \xi^{-1}(1+\xi^2\mu\nu) A_1
             \end{array} \right).
\label{tran}
\end{eqnarray}
In the following sections we will illustrate how to apply this $T$--duality
to generate the spin parameter.

\section{Einstein-Maxwell Theory}
\subsection{Spinning Electric Solution}
We start applications by generating a spinning solution from the EM
system ($\alpha=\beta=\gamma=0$) in order to demonstrate the procedure.
For a general $2+1$ EMD system with arbitrary coupling constants the
$T$--duality is identical with a linear coordinate transformation and
does not include a Harrison--type transformation.
Hence the charge parameter can not be generated by (\ref{tran}).
Thus, as the seed, one should take a charged solution.

Consider the static electrically charged BTZ solution
\cite{BaTeZa92,BaHeTeZa93} which is characterized by three parameters
$m, q, r_0$:
\begin{equation}
ds^2 = - V^2 dt^2 + r^2 d\theta^2 + V^{-2} dr^2, \qquad
A_t = -q \ln \left( \frac{r}{r_0} \right),
\end{equation}
where
\begin{equation}
V^2 = -\Lambda r^2 - m - 2 q^2 \ln \left( \frac{r}{r_0} \right).
\end{equation}
After the transformation (\ref{tran}) we get the nonvanishing metric
components
\begin{eqnarray}
h_{tt} &=&
    \xi^2 (\Lambda + \mu^2) r^2
  + \xi^2 \left[ m + 2 q^2 \ln \left( \frac{r}{r_0} \right) \right], \\
h_{t\theta} &=&
    \left[ \xi^2 \nu \Lambda + \mu (1+\xi^2 \mu \nu) \right] r^2
  + \xi^2 \nu \left[ m + 2 q^2 \ln \left( \frac{r}{r_0} \right) \right], \\
h_{\theta\theta} &=&
    \left[ \xi^2 \nu^2 \Lambda + \xi^{-2}(1+\xi^2\mu\nu)^2 \right] r^2
  + \xi^2 \nu^2 \left[ m + 2 q^2 \ln \left( \frac{r}{r_0} \right) \right].
\end{eqnarray}
To get the spin one should make a choice of parameters such that the
$r^2$ term in $h_{t\theta}$ vanish and the coefficient of the $r^2$ term
in $h_{\theta\theta}$ is equal to one.
Under this choice the three parameters $\xi, \mu$ and $\nu$ of the $SL(2,R)$
$T$--duality are related by
\begin{equation}
\xi^2 = \frac{\Lambda}{\Lambda+\mu^2}, \qquad \nu = -\frac{\mu}{\Lambda}.
\end{equation}
The remaining one is just the desired spin parameter.
After this choice the metric $h_{ij}$ becomes to
\begin{eqnarray}
h_{tt} &=&
    \Lambda r^2 + \frac{\Lambda}{\Lambda+\mu^2}
    \left[ m + 2 q^2 \ln \left( \frac{r}{r_0} \right) \right], \\
h_{t\theta} &=&
    - \frac{\mu}{\Lambda+\mu^2}
    \left[ m + 2 q^2 \ln \left( \frac{r}{r_0} \right) \right], \\
h_{\theta\theta} &=&
    r^2 + \frac{\mu^2}{\Lambda(\Lambda+\mu^2)}
  + \left[ m + 2 q^2 \ln \left( \frac{r}{r_0} \right) \right].
\end{eqnarray}
In order to have a better presentation of this solution we change the
parameters as following
\begin{equation}
\omega := \frac{\mu}{\Lambda}, \qquad
M := \frac{\Lambda}{\Lambda+\mu^2} m, \qquad
Q^2 := \frac{\Lambda}{\Lambda+\mu^2} q^2.
\end{equation}
then metric of the spinning electric solution has the usual form
\begin{eqnarray}
ds^2 =
 &-& \left[-\Lambda r^2-M-2Q^2 \ln\left( \frac{r}{r_0} \right)\right] dt^2
  - 2 \omega \left[M+2Q^2 \ln\left( \frac{r}{r_0} \right) \right] dt d\theta
     \nonumber \\
 &+& \left\{r^2+\omega^2 \left[M+2Q^2 \ln\left( \frac{r}{r_0} \right)
     \right] \right\} d\theta^2 \nonumber \\
 &+& \left\{-\Lambda r^2 - (1 + \omega^2 \Lambda)
     \left[M+2Q^2 \ln\left( \frac{r}{r_0} \right)\right] \right\}^{-1} dr^2,
 \label{sol1}
\end{eqnarray}
and the potential of the Maxwell field is
\begin{equation}
A = Q \ln \left( \frac{r}{r_0} \right) (-dt + \omega d\theta).
\end{equation}

When $M=0$, the solution (\ref{sol1}) is equivalent to the solution of
Cl\'ement \cite{Cl96}.
For the uncharged case, $Q=0$, it reduces to the BTZ solution
\cite{BaTeZa92,BaHeTeZa93} in the standard form
\begin{eqnarray}
ds^2 = &-& \left( -\Lambda R^2 - M' + \frac{J^2}{4R^2} \right) dt^2
       + R^2 \left( d\theta - \frac{J}{2R^2} dt \right)^2 \nonumber \\
       &+& \left( -\Lambda R^2 - M' + \frac{J^2}{4R^2} \right)^{-1} dr^2,
\end{eqnarray}
where
\begin{equation}
R^2 := r^2 + \omega^2 M, \quad
M'  := (1 - \omega^2 \Lambda) M, \quad
J   := 2 \omega M, \quad
\Lambda := -l^{-2}.
\end{equation}

It is worth noting that the quasilocal mass and angular momentum are
divergent (expect for the uncharged BTZ solution) because of the presence
of a logarithmic term.
This problem can be solved by introducing a Chern-Simons term \cite{Cl96}
or by a regularization \cite{KaKo97}.

\subsection{Spinning Magnetic Solution}
An electric-magnetic duality does not exist in three--dimensions.
Hence the magnetic solution should be considered separately.
In the EM theory the static magnetic monopole solution was found by
Hirschmann and Welch \cite{HiWe96}
\footnote{
     Our form of the magnetic solution is slightly different from
     \cite{HiWe96}. The original form can be achieved by the following
     transformation: $r^2 \to \frac{r^2}{l^2}-M, m \to (1+l^2) M,
     \Lambda \to -\frac1{l^2}, p \to Q_m$ and $r_0 \to 1$.
}
\begin{eqnarray}
ds^2 &=& - r^2 dt^2 + \left[ - \Lambda^{-1}r^2 + m
         + 2p^2 \ln \left(\frac{r}{r_0}\right) \right] d\theta^2 \nonumber \\
     &+& \left\{-\Lambda r^2 + \Lambda \left[ m
         + 2p^2 \ln \left(\frac{r}{r_0}\right) \right] \right\}^{-1} dr^2,
         \nonumber \\
A_\theta &=& p \ln \left( \frac{r}{r_0} \right).
\end{eqnarray}
Here we take this solution as a seed to find the corresponding spinning
solution in the same way as in the previous subsection.
After a simple calculation and a redefinition of parameters we get the
spinning magnetic solution as
\begin{eqnarray}
ds^2 =
 &-& \left\{ r^2-\omega^2\left[M + 2P^2 \ln \left( \frac{r}{r_0} \right)
     \right] \right\} dt^2
 - 2 \omega \left[M + 2P^2 \ln \left( \frac{r}{r_0} \right)
     \right] dt d\theta \nonumber \\
 &+& \left[\Lambda^{-1}r^2 + M + 2P^2 \ln \left( \frac{r}{r_0} \right)
     \right] d\theta^2 \nonumber \\
 &+& \left\{-\Lambda r^2 + \Lambda (\Lambda + \omega^2)
     \left[M + 2P^2 \ln \left( \frac{r}{r_0} \right) \right]
     \right\}^{-1} dr^2,
  \label{sol1b}
\end{eqnarray}
and
\begin{equation}
A = P \ln \left( \frac{r}{r_0} \right) (- \omega dt + d\theta).
\end{equation}

\subsection{Static Dyon}
Furthermore, both spinning electric or magnetic solutions can be treated
as a static dyon by setting $\omega = P/Q$ in the electric
or $\omega = Q/P$ in the magnetic solutions.
The static dyon is
\begin{eqnarray}
ds^2 =
 &-& \left[-\Lambda r^2-M-2Q^2 \ln\left( \frac{r}{r_0} \right) \right] dt^2
 - 2 \frac{P}{Q}\left[M+2Q^2 \ln\left( \frac{r}{r_0} \right)\right] dt d\theta
     \nonumber \\
 &+& \left\{r^2 + \frac{P^2}{Q^2} \left[M+2Q^2\ln\left(\frac{r}{r_0}\right)
     \right] \right\} d\theta^2 \nonumber \\
 &+& \left\{-\Lambda r^2 - (1 + \frac{P^2}{Q^2} \Lambda)
     \left[M+2Q^2 \ln\left( \frac{r}{r_0} \right) \right] \right\}^{-1} dr^2,
\end{eqnarray}
with
\begin{equation}
A = \ln \left( \frac{r}{r_0} \right) (- Q dt + P d\theta).
\end{equation}
This is a generalization of the (anti-) self dual solutions
\cite{KaKo95} which reduces to them if $Q=\pm P$.

\section{Brans-Dicke Theory}
In the string frame, a set of static and stationary solutions of
Brans--Dicke theory, $\beta=4$ and $F=0$, was found in
\cite{SaKlLe96,SaLe98}.
Here we consider similar solutions in the Einstein frame.
The static solutions can be easy obtained by solving the field equations
under Schwarzschild gauge, i.e. $-g_{tt}=g_{rr}^{-1}$.
Then the corresponding stationary solution can be found by $T$--duality.

According to the coupling constant $\gamma$, the static solutions can be
divided into four case.
For one of them, $\gamma=0$, there exist only asymptotically flat
solutions, i.e. $\Lambda$ should be zero. So it is not considered here.

\subsection{Case $\gamma \ne -1, 0, \frac12$}
The static solution for the case $\gamma \ne -1, 0, \frac12$, by solving
field equations, is
\begin{eqnarray}
ds^2 &=& - W_1(r) dt^2 + c^2 r^{\frac{2\gamma}{\gamma+1}} d\theta^2
         + W_1^{-1}(r) dr^2, \nonumber \\
\phi &=& -\frac1{2(\gamma+1)} \ln\left( \frac{r}{r_0} \right),
\end{eqnarray}
where
\begin{equation}
W_1(r) = -a r^{\frac{2\gamma}{\gamma+1}} - m r^{\frac1{\gamma+1}}, \qquad
a := \frac{2(\gamma+1)^2\Lambda}{\gamma(2\gamma-1)} r_0^{\frac2{\gamma+1}}.
\end{equation}
This solution is related to the solution of \cite{SaKlLe96}.
\footnote{
     The relationship is: $\gamma \to \omega + 2,
     r \to r^\frac{\omega+3}{\omega+1},
     r_0 \to a^{-\frac{\omega+3}{\omega+1}},
     c^2 \to \frac{\omega+3}{\omega+1},
     m \to \frac{\omega+3}{\omega+1}b a^\frac{-1}{\omega+1},
     \Lambda \to -2 \lambda^2.$
}
Using it as a seed we apply the $T$--duality and get
\begin{eqnarray}
h_{tt} &=&
     \xi^2 (a + \mu^2 c^2) r^{\frac{2\gamma}{\gamma+1}}
   + \xi^2 m r^\frac1{\gamma+1}, \\
h_{t\theta} &=&
     \left[\xi^2\nu a+\mu(1+\xi^2\mu\nu)c^2\right]r^{\frac{2\gamma}{\gamma+1}}
   + \xi^2\nu m r^\frac1{\gamma+1}, \\
h_{\theta\theta} &=&
     \left[\xi^2\nu^2 a+\xi^{-2}(1+\xi^2\mu\nu)^2c^2\right]
     r^{\frac{2\gamma}{\gamma+1}} + \xi^2\nu^2 m r^\frac1{\gamma+1}.
\end{eqnarray}
In this case, in order to generate the spin we should choose the parameters
in such a way that the $r^2$ term in $h_{t\theta}$ vanishes
and the coefficient of $r^2$ term in $h_{\theta\theta}$ is equal to $c^2$.
This corresponds to the following relations
\begin{equation}
\xi^2 = \frac{a}{a+\mu^2 c^2}, \qquad \nu = -\frac{\mu c^2}{a},
\end{equation}
and the metric $h_{ij}$ reduces to
\begin{eqnarray}
h_{tt} &=&
     a r^{\frac{2\gamma}{\gamma+1}}
   + \frac{a}{a+\mu^2 c^2} m r^\frac1{\gamma+1}, \\
h_{t\theta} &=&
   - \frac{\mu c^2}{a+\mu^2 c^2} m r^\frac1{\gamma+1}, \\
h_{\theta\theta} &=&
     c^2 r^{\frac{2\gamma}{\gamma+1}}
   + \frac{\mu^2 c^4}{a(a+\mu^2 c^2)} m r^\frac1{\gamma+1}.
\end{eqnarray}
Redefining the parameters as
\begin{eqnarray}
\omega &:=& \frac{\mu c^2}{a}, \qquad
  M := \frac{a}{a+\mu^2 c^2} m,
\end{eqnarray}
one obtains the metric of new solution as
\begin{eqnarray}
ds^2 =
  &-& \left(- a r^{\frac{2\gamma}{\gamma+1}} - M r^\frac1{\gamma+1}
      \right) dt^2 - 2 \omega M r^\frac1{\gamma+1} dt d\theta \nonumber \\
  &+& \left( c^2 r^{\frac{2\gamma}{\gamma+1}} + \omega^2 M r^\frac1{\gamma+1}
    \right) d\theta^2 \nonumber \\
  &+& \left[- a r^{\frac{2\gamma}{\gamma+1}}
   - (1+\frac{\omega^2}{c^2}a) M r^\frac1{\gamma+1} \right]^{-1} dr^2.
\end{eqnarray}
The dilaton is invariant under the transformation.

\subsection{Case $\gamma = \frac12$}
The static solution for the case $\gamma = \frac12$ is
\begin{eqnarray}
ds^2 &=& - W_2(r) dt^2 + c^2 r^\frac23 d\theta^2
         + W_2^{-1}(r) dr^2, \nonumber \\
\phi &=& -\frac13 \ln\left( \frac{r}{r_0} \right),
\end{eqnarray}
where
\begin{equation}
W_2(r) = - \frac32 \Lambda r_0^\frac43 r^\frac23 \ln(br).
\end{equation}
After $T$--duality the metric becomes
\begin{eqnarray}
h_{tt} &=&
   - \xi^2 W_2(r) + \xi^2 \mu^2 c^2 r^\frac23, \\
h_{t\theta} &=&
   - \xi^2 \nu W_2(r) + \mu(1+\xi^2\mu\nu) c^2 r^\frac23, \\
h_{\theta\theta} &=&
   - \xi^2\nu^2 W_2(r) + \xi^{-2}(1+\xi^2\mu\nu)^2 c^2 r^\frac23.
\end{eqnarray}

There are two possibilities of stationary solutions depending on the choice
of parameters. One choice is
\begin{equation}
\xi^2 = 1, \qquad \mu = 0, \qquad \nu = \omega,
\end{equation}
and the stationary solution takes the form
\begin{eqnarray}
ds^2 = &-& W_2(r) dt^2 - 2 \omega W_2(r) dt d\theta
    + \left[ c^2 r^\frac23 - \omega^2 W_2(r) \right] d\theta^2 \nonumber \\
  &+& W_2^{-1}(r) dr^2.
\end{eqnarray}
The other choice is
\begin{equation}
\xi^2 = 1, \qquad \mu = -\omega, \qquad \nu = 0,
\end{equation}
then the stationary solution becomes
\begin{eqnarray}
ds^2 &=& \left[- W_2(r) + \omega^2 c^2 r^\frac23 \right] dt^2
  - 2 \omega c^2 r^\frac23 dt d\theta + c^2 r^\frac23 d\theta^2 \nonumber \\
  &+& W_2^{-1}(r) dr^2.
\end{eqnarray}

\subsection{Case $\gamma = -1$}
For the case $\gamma = -1$ the Schwarzschild gauge does not hold.
However, the static solution is not difficult to obtain
\begin{eqnarray}
ds^2 &=& -b^2 r^2 dt^2 + c^2 r^2 d\theta^2
   + \left( -\frac23 \Lambda r_0^{-2} r^4 \right)^{-1} dr^2, \nonumber \\
\phi &=& -\frac12 \ln\left( \frac{r}{r_0} \right),
\end{eqnarray}

After $T$--duality the metric becomes
\begin{eqnarray}
h_{tt} &=&
   - \xi^2 b^2 r^2 + \xi^2 \mu^2 c^2 r^2, \\
h_{t\theta} &=&
   - \xi^2 \nu b^2 r^2 + \mu(1+\xi^2\mu\nu) c^2 r^2, \\
h_{\theta\theta} &=&
   - \xi^2\nu^2 b^2 r^2 + \xi^{-2}(1+\xi^2\mu\nu)^2 c^2 r^2.
\end{eqnarray}
We make the following choice of the parameters
\begin{equation}
\xi^2 = 1, \qquad \mu = 0, \qquad \nu = \omega.
\end{equation}
The stationary solution is
\begin{eqnarray}
ds^2 = &-& b^2 r^2 dt^2 - 2 \omega b^2 r^2 dt d\theta
   + \left( c^2 r^2 - \omega^2 b^2 r^2 \right) d\theta^2 \nonumber \\
   &+& \left( -\frac23 \Lambda r_0^{-2} r^4 \right)^{-1} dr^2.
\end{eqnarray}

\section{Einstein-Maxwell-Dilaton Theory}
\subsection{Spinning Electric Solution}
A general static electric charged solution in the EMD theory was found by
Chan and Mann \cite{ChMa94}.
It is characterized by four constants $c, m, q, r_0$
\begin{eqnarray}
ds^2 &=& -U(r) dt^2 + c^2 r^N d\theta^2 + U^{-1}(r) dr^2, \nonumber \\
\phi &=& k \ln \left( \frac{r}{r_0} \right), \qquad
A_t = -q r^{\frac{N}{2}-1},
\end{eqnarray}
where
\begin{eqnarray}
U(r) &=& - \frac{8\Lambda r_0^{2-N}}{N(3N-2)} r^N - m r^{1-\frac{N}{2}}
       - \frac{2(N-2)r_0^{N-2}}{N} q^2, \nonumber \\
k &=& \pm \frac1{4} \sqrt{\frac{N(2-N)}{\gamma}}, \nonumber \\
4 \alpha k &=& \beta k = N - 2, \qquad 4\alpha = \beta, \qquad 0<N<2.
\end{eqnarray}
Taking this solution as a seed, the corresponding $H_0$ and $H_1$ in
(\ref{seed}) are
\begin{equation}
H_0^2 = -B r^N - m r^{1-\frac{N}{2}} - C, \qquad H_1^2 = c^2 r^N.
\end{equation}
Here we introduce following combinations of the parameters
\begin{equation}
B := \frac{8 \Lambda r_0^{2-N}}{N(3N-2)}, \qquad
C := \frac{2(N-2)r_0^{N-2}}{N} q^2,
\end{equation}
to simplify the notation.
After the $T$--duality transformation we have
\begin{eqnarray}
h_{tt} &=&
     \xi^2 (B + \mu^2 c^2) r^N
   + \xi^2 \left( m r^{1-\frac{N}{2}} + C \right), \\
h_{t\theta} &=&
     \left[ \xi^2 \nu B + \mu (1+\xi^2\mu\nu) c^2 \right] r^N
   + \xi^2 \nu \left( m r^{1-\frac{N}{2}} + C \right), \\
h_{\theta\theta} &=&
     \left[ \xi^2 \nu^2 B + \xi^{-2}(1+\xi^2\mu\nu)^2 c^2 \right] r^N
   + \xi^2 \nu^2 \left( m r^{1-\frac{N}{2}} + C \right).
\end{eqnarray}
We should make a choice of parameters such that
the $r^2$ term in $h_{t\theta}$ vanishes and the coefficient of $r^2$
term in $h_{\theta\theta}$ is equal to $c^2$, which means
\begin{equation}
\xi^2 = \frac{B}{B+\mu^2 c^2}, \qquad \nu = -\frac{\mu c^2}{B}.
\end{equation}
Then the metric $h_{ij}$ becomes
\begin{eqnarray}
h_{tt} &=&
  B r^N + \frac{B}{B+\mu^2 c^2} \left( m r^{1-\frac{N}{2}} + C \right), \\
h_{t\theta} &=&
   - \frac{\mu c^2}{B+\mu^2 c^2} \left( m r^{1-\frac{N}{2}} + C \right), \\
h_{\theta\theta} &=&
     c^2 r^N
   + \frac{\mu^2 c^4}{B(B+\mu^2 c^2)} \left( m r^{1-\frac{N}{2}} + C \right).
\end{eqnarray}
It is convenient to present the spinning charged solution in terms
of the following variables
\begin{eqnarray}
\omega &:=& \frac{\mu c^2}{B}, \qquad
  M := \frac{B}{B+\mu^2 c^2} m, \qquad
  Q^2 := \frac{B}{B+\mu^2 c^2} q^2, \nonumber \\
C' &:=& \frac{2(N-2)r_0^{N-2}}{N} Q^2.
\end{eqnarray}
Then the metric takes the form
\begin{eqnarray}
ds^2 =
  &-& \left(- B r^N - M r^{1-\frac{N}{2}} - C' \right) dt^2
  - 2 \omega \left( M r^{1-\frac{N}{2}} + C'\right) dt d\theta
    \nonumber \\
  &+& \left[ c^2 r^N + \omega^2 \left( M r^{1-\frac{N}{2}} + C' \right)
    \right] d\theta^2 \nonumber \\
  &+& \left[- B r^N - (1+\frac{\omega^2}{c^2}B)(M r^{1-\frac{N}{2}} + C')
      \right]^{-1} dr^2.
\end{eqnarray}
The dilaton and the $1$--form are
\begin{equation}
\phi = k \ln \left( \frac{r}{r_0} \right), \qquad
A = Q r^{\frac{N}{2} -1 } (-dt + \omega d\theta).
\end{equation}
For the uncharged case, $q=0$, our solution reduces to the spinning black
hole solution in \cite{ChMa96} with the parameter normalization
$r_0^{2-N}=c^2$.

The horizon radius is defined by the equation
\begin{equation}
B r_H^N + (1+\omega^2 \frac{B}{c^2})(M r_H^{1-\frac{N}{2}} + C') = 0.
\end{equation}
Unfortunately it can not be solved explicitly.
But it is easy to check that the charged spinning black holes could have
an inner horizon.
Moreover the horizon in the extremal case can be found in general
\begin{equation}
r_H^{\hbox{(ext)}} =
  \left[ \frac{(2-N)(c^2+\omega^2 B)M}{2c^2NB} \right]^\frac{2}{3N-2},
\end{equation}
and the extremality condition is
\begin{eqnarray}
&& B^\frac{N-2}{3N-2}\left[(1+\frac{\omega^2 B}{c^2})M\right]^\frac{2}{3N-2}
  \left\{\left( \frac{2-N}{2N} \right)^\frac{2N}{3N-2}
  - \left( \frac{2-N}{2N} \right)^\frac{2-N}{3N-2} \right\}
  \nonumber \\
&& + \left( 1+\frac{\omega^2 B}{c^2} \right) C' = 0.
\end{eqnarray}

Besides this, let us consider the $g_{\theta\theta}$ component of the metric
\begin{equation}
g_{\theta\theta} = c^2 r^N+\omega^2 \left( M r^{1-\frac{N}{2}}+ C' \right).
\end{equation}
In general we should assume this term to be positive otherwise the
angle $\theta$ will becomes a time--like coordinate which will produce
a closed time--like curve.
This requirement always holds for the static solution and can be easy achieved
by assuming $M>0$ for spinning uncharged solution.
But due to the term $C'$ it could be negative so that we can not preclude the
existence of a closed time--like curve for the spinning charged solution.

\subsection{Spinning Magnetic Solution}
Here we consider separately the generation of a spinning magnetic charged
solution in the EMD theory.
It is not difficult, by solving field equations like those in \cite{ChMa94},
to find the following static magnetic solution
\begin{eqnarray}
ds^2 = && - c^2 r^N dt^2
          + \left(-B r^N + m r^{1+\frac{N}{2}} + C \right) d\theta^2
          \nonumber \\
       && + \left(-B r^N + m r^{1+\frac{N}{2}} + C \right)^{-1} dr^2,
          \nonumber \\
\phi = && k \ln \left( \frac{r}{r_0} \right), \qquad
   A_\theta = p r^{\frac{N}{2}-1}.
\end{eqnarray}
Using this solution as a seed, after the $SL(2,R)$ transformation
(\ref{tran}) and a suitable choice of parameters, we get a spinning magnetic
solution. The metric of this new solution is
\begin{eqnarray}
ds^2 = &-& \left[ c r^2-\omega^2 \left( M r^{1-\frac{N}{2}}+C' \right)
           \right] dt^2
       - 2 \omega \left( M r^{1-\frac{N}{2}} + C' \right) dt d\theta
       \nonumber \\
       &+& \left( -B r^N + M r^{1-\frac{N}{2}} + C' \right) d\theta^2
         \nonumber \\
       &+& \left[ -B r^N + \left( 1+\frac{\omega^2}{c^2}B \right)
           \left( M r^{1-\frac{N}{2}} + C' \right) \right]^{-1} dr^2,
\end{eqnarray}
and the dilaton and the form fields are
\begin{equation}
\phi = k \ln \left( \frac{r}{r_0} \right), \qquad
A = P r^{\frac{N}{2} -1 } (- \omega dt + d\theta).
\end{equation}

\subsection{Static Dyon Solution}
Both spinning electric or magnetic solutions can be treated as a dyon
solution by setting $\omega = P/Q$ in the electric or $\omega = Q/P$ in
the magnetic solutions.
The dyon solution is
\begin{eqnarray}
ds^2 =
  &-& \left(- B r^N - M r^{1-\frac{N}{2}} - C' \right) dt^2
  - 2 \frac{P}{Q} \left( M r^{1-\frac{N}{2}} + C'\right) dt d\theta
    \nonumber \\
  &+& \left[ c^2 r^N + \frac{P^2}{Q^2} \left( M r^{1-\frac{N}{2}} + C' \right)
    \right] d\theta^2 \nonumber \\
  &+& \left[- B r^N - (1+\frac{P^2}{c^2Q^2}B)(M r^{1-\frac{N}{2}} + C')
      \right]^{-1} dr^2,
\end{eqnarray}
and
\begin{equation}
\phi = k \ln \left( \frac{r}{r_0} \right), \qquad
A = r^{\frac{N}{2} -1 } (- Q dt + P d\theta).
\end{equation}

\section{Conclusion}
Using the $T$--duality symmetry in $2+1$--dimensional gravity theory we
can generate the spin parameter.
It enables us to find spinning charged solutions from static ones without
solving the complicated field equations.
Applying this symmetry we recover the spinning electric charged solution
\cite{Cl96} and obtain a new spinning magnetically charged solution and
a static dyon in the Einstein-Maxwell theory with a cosmological constant.

For the Brans-Dicke theory a set of static and stationary solutions
was obtained in the Einstein frame. Most of them are equivalent the
solution of \cite{SaKlLe96,SaLe98}.
For the EMD theory we find new spinning electric and magnetic charged
solutions.
The existence of a charge provides the possibility of an inner horizon.
The static dyon solution is also presented.
All the new spinning charged solutions have quite different properties
>from the earlier static charged or spinning uncharged solutions.
This point requires further investigation.

The solutions presented here do not seem to be the most general ones
with or without a dilaton field.
A general $SL(2,R)$ transformation includes three parameters.
But by a choice of parameters we had restricted them to produce only one
spin parameter.
It seems that this condition can be relaxed.
One of the other two parameters corresponds to rescaling of the radial
coordinate, but the other which produces an $r^2$ term in $N^\theta$ can
lead to a globally different structure.
It is equivalent to applying a coordinate transformation
$\theta\to\theta+T t$ to the solutions derived here.

The BTZ solution can be interpreted as an exact three-dimensional black
string solution \cite{HoWe93}.
In three dimensions the $3$--form field is equivalent to a cosmological
constant.
So the solutions presented here should have the corresponding charged
string counterparts.

\section*{Acknowledgments}
It is a pleasure to thank D. V. Gal'tsov and J. M. Nester for helpful
discussions and suggestions.

\end{document}